\documentclass[12pt]{article}
\usepackage{graphicx}

\newcommand\pubnumber{SNSN-323-63}
\newcommand\pubdate{\today}
\usepackage{url}
\def\frascati{Laboratori Nazionali di Frascati dell'INFN\\
Via E. Fermi, 40, I-00044 Frascati, ITALY}

\def\nc{North Carolina State University,\\ Department of Physics,\\
Raleigh, NC, USA 27695} 
\usepackage{xspace,cite}

\newcommand{\ke}{\ensuremath{K \to e \nu}\xspace }
\newcommand{\km}{\ensuremath{K \to \mu \nu}\xspace}

\newcommand{\cO}{\mathcal{O}}

\newcommand{\kedg}{\ensuremath{K_{e2\gamma}}\xspace}

\def\Title#1{\begin{center} {\Large #1 } \end{center}}
\def\Author#1{\begin{center}{ \sc #1} \end{center}}
\def\Address#1{\begin{center}{ \it #1} \end{center}}

\newcommand\pubblock{\rightline{\begin{tabular}{l} \pubnumber\\
         \pubdate  \end{tabular}}}
\newenvironment{Abstract}{\begin{quotation}  }{\end{quotation}}
\newenvironment{Presented}{\begin{quotation} \begin{center} 
             PRESENTED AT\end{center}\bigskip 
      \begin{center}\begin{large}}{\end{large}\end{center} \end{quotation}}

\def\beq{\begin{equation}}
\def\eeq#1{\label{#1}\end{equation}}
\def\eeqn{\end{equation}}

\def\beqa{\begin{eqnarray}}
\def\eeqa#1{\label{#1}\end{eqnarray}}
\def\eeqan{\end{eqnarray}}

\let\bar=\overbar

\def\Dslash{\not{\hbox{\kern-4pt $D$}}}
\def\dslash{\not{\hbox{\kern-2pt $\del$}}}

\def\msb{{\bar{\ssstyle M \kern -1pt S}}}

\begin{document}
\begin{titlepage}
\pubblock

\vfill
\Title{$V_{ud}$ and $V_{us}$: working group I summary}
\vfill
\Author{Tommaso Spadaro\Address{\frascati}}
\Author{Albert R. Young\Address{\nc}}

\vfill
\begin{Abstract}
The present status of universality tests of the weak couplings for quarks and leptons is reviewed, with updated
information for the $\mu$ lifetime and for first-row inputs in the CKM matrix. 
We discuss the impact of this high-precision SM test in constraining new physics models. We
also discuss a precise lepton flavor-violation test from leptonic $K$
decays and recent progress in $K$--$\bar K$ mixing.
\end{Abstract}
\vfill
\begin{Presented}
CKM workshop\\
Warwick, UK,  September 6--10, 2010
\end{Presented}
\vfill
\end{titlepage}
\def\thefootnote{\fnsymbol{footnote}}
\setcounter{footnote}{0}

\section{Testing new physics from tree-level mediated decays}
In the search for new physics (NP) effects beyond the Standard Model (SM), tests of the
universality of the weak coupling between leptons and quarks are particularly relevant, in that
sensitivities to mass scales at the TeV level and beyond can be reached for a wide range of extensions
to the standard model. The door to these flavor universality tests is the precise measurement of
the muon lifetime, $\tau_\mu$, from which the Fermi constant for leptonic weak interactions, $G_\mu$,
is extracted: 
\begin{equation}
\frac{1}{\tau_\mu}=\frac{G_\mu^2m_\mu^5}{192\pi^3}\left(1 + \Delta q\right),
\end{equation}
where $\Delta q$ represents QED radiative corrections and small phase space effects\cite{PDG2010}.
The results of an experiment with roughly an order of magnitude improvement
for the precision of $\tau_{\mu}$ were presented by P. Debevec for the MuLan collaboration~\cite{debevec-CKM2010}.

\begin{figure}
\centering
\includegraphics[height=6.0cm]{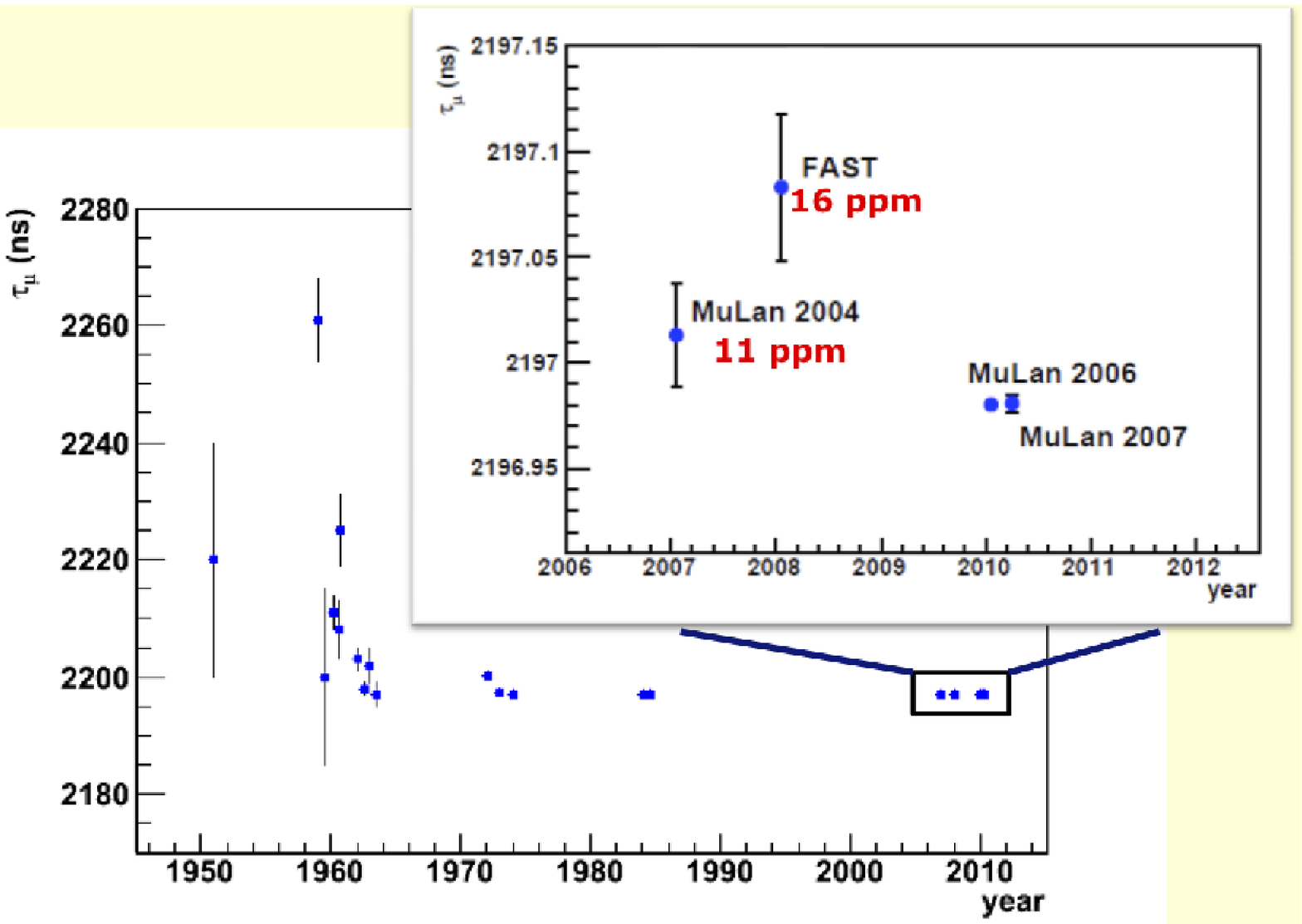} 
\includegraphics[height=6.0cm]{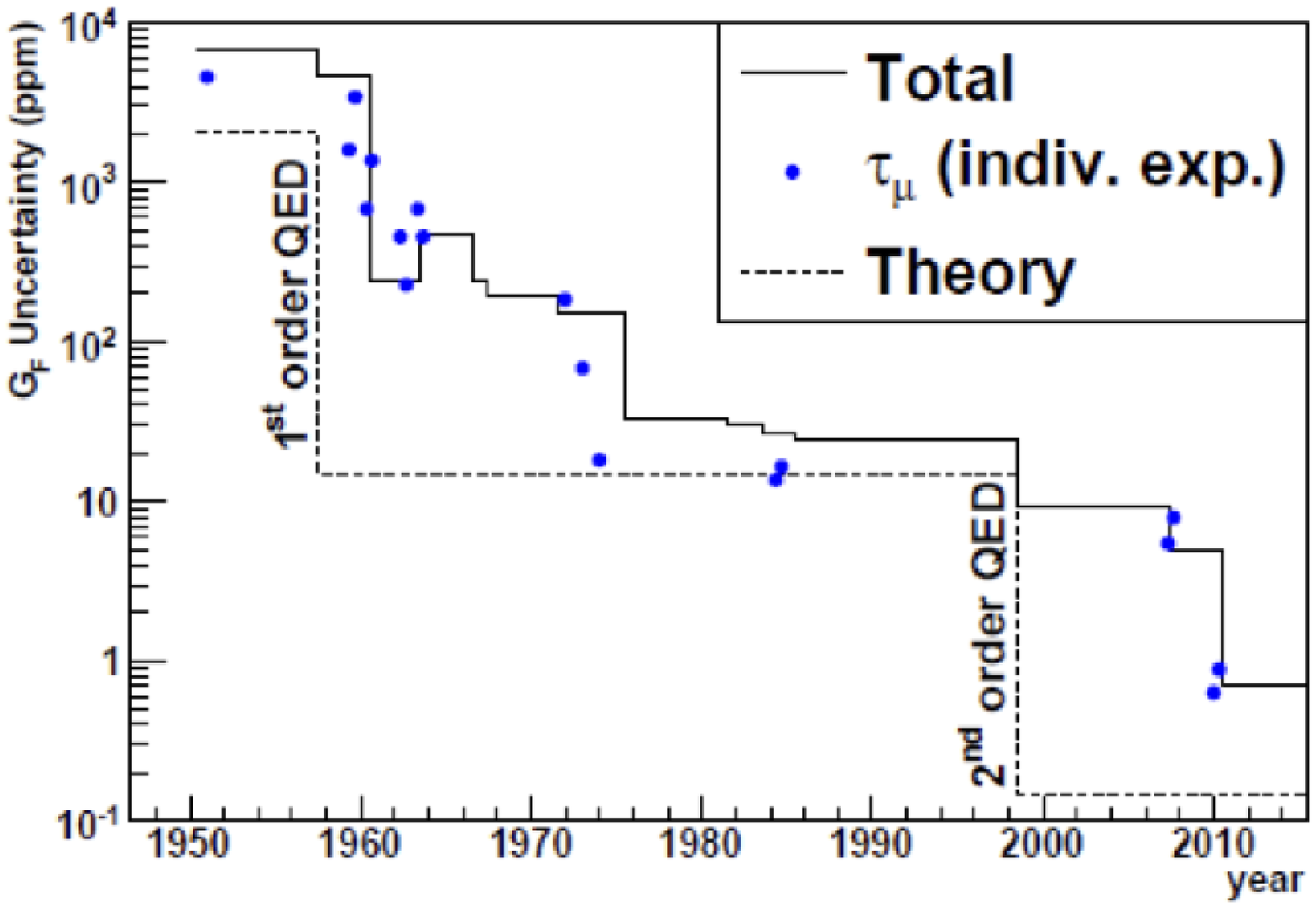}
\caption{Figures from~\cite{debevec-CKM2010}. Left: measurements of the muon lifetime from 1950 to the present. Right:
experimental and theoretical uncertainty on $G_\mu$ from 1950 to the present.}
\label{fig:gmu}
\end{figure}

In order to achieve this improvement, the MuLan approach was to use the high intensity surface muon
beams at PSI, together with a fast kicker to define accumulation and counting periods.  They characterized
the effects of one of the key sources of potential systematic error, muon spin precession and relaxation,
using two different stopping targets: in 2006, a ferromagnetic Arnokrome$\textsuperscript{\textregistered}$\ target, and
in 2007, a quartz target immersed in a nearly uniform 130G field.  In the first case, the high internal
field ensured the effective polarization of the muon sample was reduced by about three orders of magnitude,
in the second case the experiment monitored the precession (with a period of roughly 550 ns) of roughly 10\%
of the sample.  In addition, detectors were arranged symmetrically around the stopping target (and covering 75\% of $4\pi$)
to ensure very low sensitivity to spin precession and relaxation.  
Ultimately, the uncertainty
of muon lifetimes determined by the MuLan collaboration were dominated by statistical errors, with a
lifetime result of $\tau$(MuLan)$=2196980.3(2.2)$ps (1.0 ppm). The results for the two different MuLan
targets are in impressive agreement with previous $\tau _{\mu}$ (Fig.~\ref{fig:gmu}, left). To the reviewers'
knowledge, this is the most precise lifetime measurement ever performed.

The MuLan measurement pushes the current experimental sensitivity to the same order of magnitude as the
precision of available theoretical treatments of muon decay.  Radiative corrections including
virtual one-loop diagrams and inclusive bremsstrahlung amount to $\sim-4\times10^{-3}$ and are known to
better than a part in $10^{-6}$~\cite{marciano}, and have been refined as recently as 2008\cite{Pak08}. After 60 years
of experimental and theoretical development, the total uncertainty on $G_\mu$ is at the $10^{-6}$ level
(Fig.~\ref{fig:gmu}, right). 
The value recommended by the MuLan collaboration for the effective coupling constant $G_{\mu}$ is
\begin{equation}
G_\mu = 1.1663788(7)\times10^{-5}~\mathrm{GeV}^{-2},
\end{equation}
and is the reference benchmark for the Fermi coupling, $G_F \equiv G_{\mu}$. 
Given the standard model assumptions of CKM unitarity and the equivalence of quark and lepton weak couplings,
this value of the Fermi coupling can be compared with
\begin{equation}
G_\mathrm{quarks}^2 = G_F^2 \left|V_{ud}\right|^2 + G_F^2 \left|V_{us}\right|^2,
\end{equation}
which is at present best known from the decay widths of superallowed Fermi transitions of specific nuclei
(Sec.~\ref{sec:vud}) and of leptonic and semileptonic decays of kaons (Sec.~\ref{sec:vus}). We present this comparison
in Sec. ~\ref{sec:end}. 
In the last equation, the contribution of $|V_{ub}|^2$ can be safely neglected. 

\section{Status of $V_{ud}$}
\label{sec:vud}
High precision values of the constant $V_{ud}$ can be extracted from measurements of the superallowed
nuclear decays, decays in mirror nuclei (including the neutron), and pion decay.  The status of the
measurement programs for each of these systems is reviewed in the following subsections. 
\subsection{Superallowed $0^+\rightarrow 0^+ \beta$-decay} 
\begin{figure}
\centering
\includegraphics[height=6.0 cm]{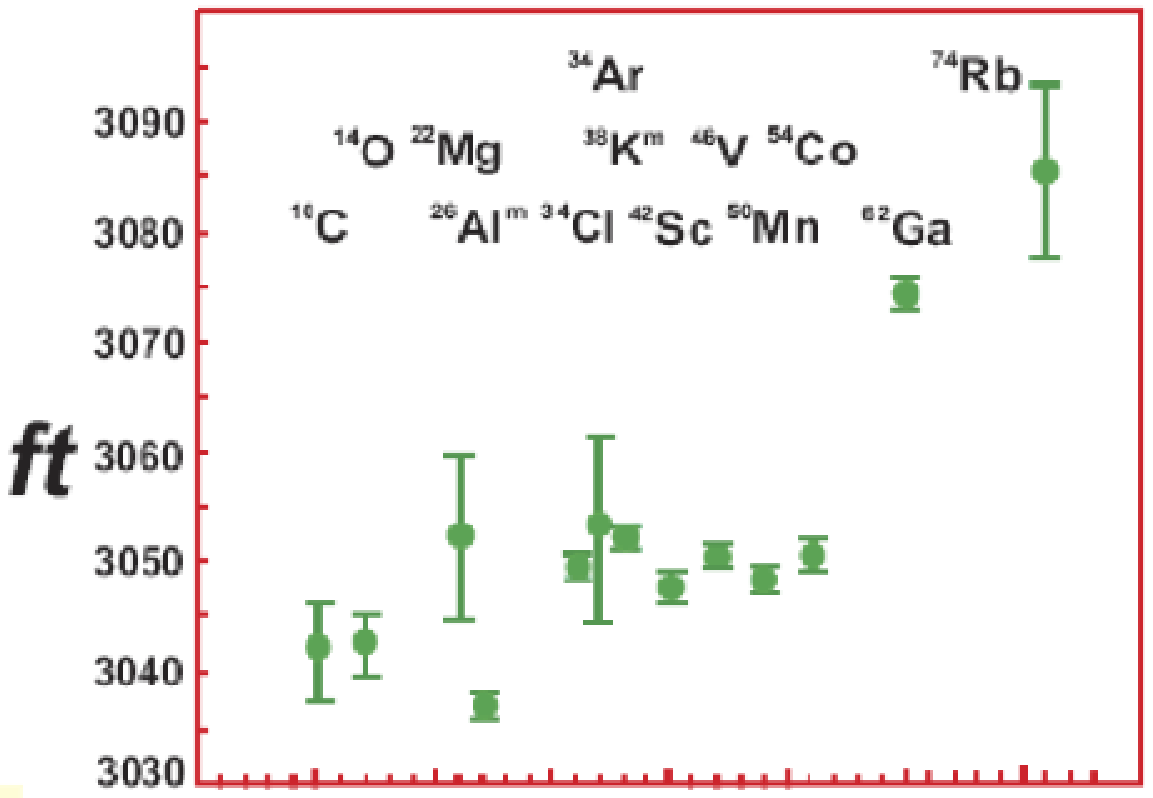}
\includegraphics[height=6.0 cm]{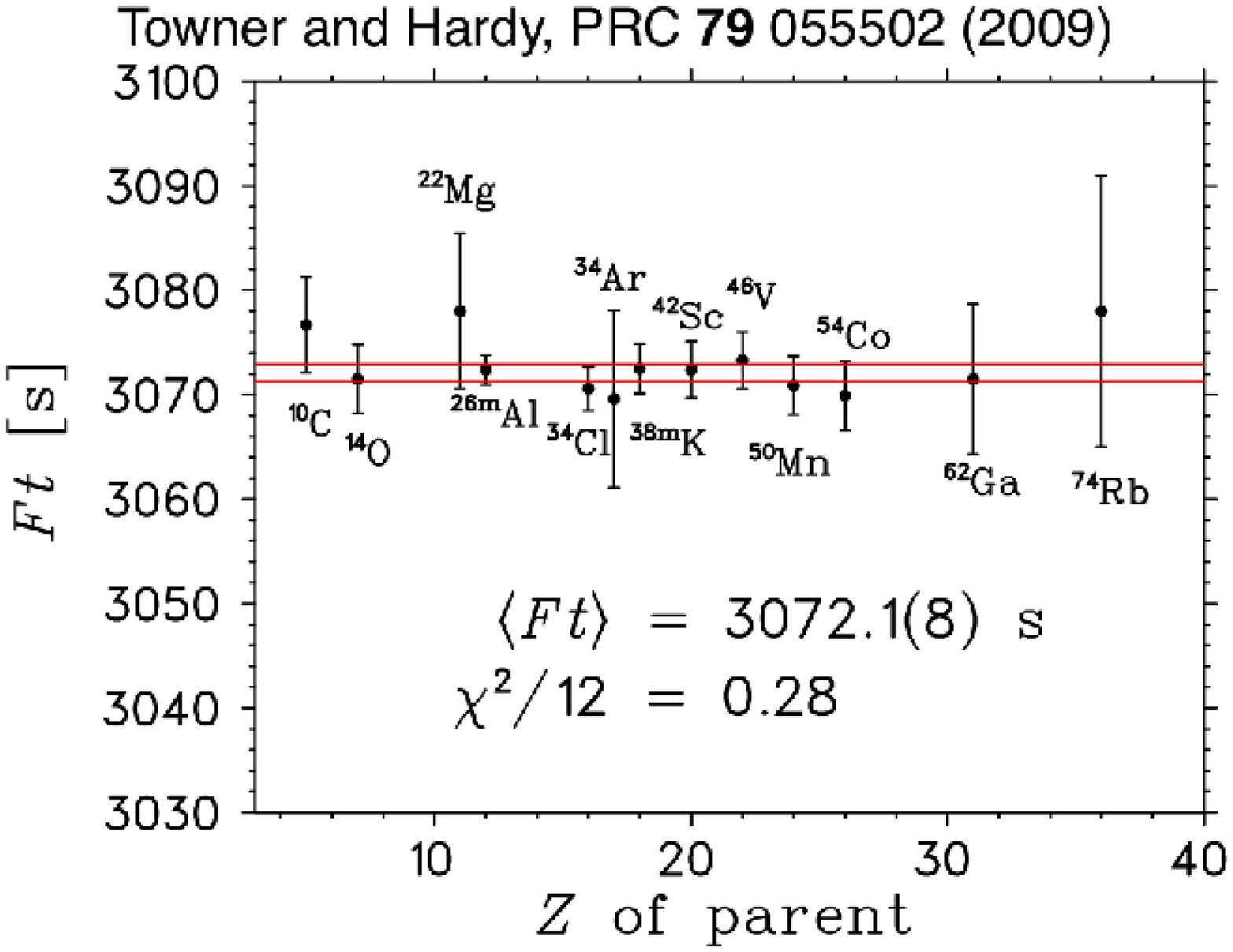}
\caption{Pictures from~\cite{TownerHardy09}. 
Left: uncorrected $ft$ values for different $0^+\to 0^+$ $T=1$ nuclear $\beta$ decays. Right: corrected $Ft$ values.}
\label{fig:Ft}
\end{figure}

At present, the most precise determination of $V_{ud}$ comes from $0^+\to 0^+$ vector transitions of
specific isospin-one ($I=1$) nuclei. These transitions benefit from the conservation of the vector current and
from small isospin-breaking corrections. The determination of $V_{ud}$ involves three measurements: the
energy separation, $Q_{EC}$, between initial and final nuclear states (required for the calculation of the
transition phase space, $f$); the half life $t_{1/2}$ of the parent nucleus and the branching ratio
$\mathrm{BR}(0^+\to0^+)$ for transitions between isobaric analog states.  These inputs are combined
in a quantity which should be nucleus-independent to first order:
\begin{equation}
ft \equiv f\times t_{1/2}\times \mathrm{BR} = \frac{K}{2G_F^2\left|V_{ud}\right|^2},
\end{equation}
where $K/(\hbar c)^6=8120.2787(11)\times 10^{-10}$ GeV$^{-4}$s\cite{TownerHardy09}. Values of $ft$ are plotted
in the left panel of Fig.~\ref{fig:Ft},
showing an evident residual nucleus dependence. The above picture has indeed to be corrected accounting for the fact
that the decay occurs within the nuclear medium. The following quantity is expected to be nucleus-independent: 
\begin{equation}
Ft \equiv f\times t_{1/2}\times \mathrm{BR}\times\left(1+\delta^\prime_R\right)\left(1+\delta_R\right) = 
\frac{K}{2G_F^2\left|V_{ud}\right|^2\left|M_F\right|^2\left(1+\Delta_R^V\right)},
\end{equation}
where $M_F$ is the Fermi strength for the decay and is equal to $\sqrt{2}$ for pure Fermi, $T=1$ transitions.
Radiative processes induce various contributions. The first contribution, $\delta^\prime_R\sim1.5\%$,
depends both on the atomic number $Z$ and on the maximum energy $E_\mathrm{max}$ of $\beta$-decay electron. A
second term, $\delta_R\sim0.5\%,$ includes nuclear-dependent radiative corrections and isospin symmetry-breaking
corrections. A further correction, $\Delta_R^V\sim2.4\%$, modifies the effective coupling and is a
transition-independent radiative correction. The plot in the right panel of Fig.~\ref{fig:Ft} shows that corrected
$Ft$ values are independent of $Z$ to within three parts in $10^4$.

Over 200 measurements contribute to the corrected $Ft$ values for superallowed $\beta$ decays of 13 nuclei. Given
the huge body of high precision data and the fact that the limiting uncertainty to $V_{ud}$ comes from theoretical
corrections to the $Ft$ values, these measurements are expected to produce modest incremental improvement in the
precision of $V_{ud}$.  As reported by Melconian\cite{melconian-CKM2010}, recent effort has focused primarily on
remeasurement of the
$Q_{EC}$ values for a large number of decays, resulting in significant shifts in the values for $^{42}$Sc, $^{46}$V,
$^{50}$Mn and $^{54}$Co, and revised (but not signficantly changed) values for 13 other species. Half-life measurements
for $^{26}$Al$^{m}$ (mirror nucleus, see below), 
$^{29}$P, $^{31}$S (also its branching ratio), $^{39}$Ca, $^{10}$C, and $^{26}$Si are also being
carried out.


After an update in 2009~\cite{TownerHardy09} the result for $V_{ud}$ was
\begin{equation}
\label{eq:vudsa}
V_{ud} = 0.97425(08)_\mathrm{exp}(21)_\mathrm{theory},
\end{equation}
where the first error accounts for the uncertainty in the experimental data, while the second error refers to the
theoretical corrections, dominated by the transition-independent correction $\Delta_R^V$, with an uncertainty which
contributes $\pm 0.00018$ to the error budget.  As reported by Marciano\cite{marciano-CKM2010} the
magnitude of the radiative corrections is stable and unchallenged at present.  The uncertainty has also been very stable,
with roughly a factor of two improvement in 2006\cite{MarcianoSirlin06}, still leaving this the dominant uncertainty for
$\beta$-decay studies in the $u$-$d$ quark sector.

Although they contribute less to the uncertainty budget than $\Delta_R^V$, the nuclear structure-dependent
corrections continue to attract attention. As reviewed by Towner in these proceedings\cite{towner-CKM2010}, a
number of alternate methods for handling these corrections have been proposed, but only the shell model calculations
using Saxon-Woods radial wavefunctions of Towner and Hardy predict the $Z$ dependence for the isospin-mixing
corrections expected from the Conserved Vector Current (CVC) hypothesis. Given the subtlety and importance of nuclear
structure effects (particularly the isospin-violating corrections), there is a general recognition that a
robust cross-check of these calculations is required.  Several strategies are being pursued\cite{melconian-CKM2010}.
Within the $I=1$ systems, higher mass transitions, such as $^{62}$Ga and $^{74}$Rb are being measured to provide
stringent tests of the isopsin-mixing corrections. Another strategy to test isospin-mixing models is to
perform high precision measurements in superallowed transitions with isospin other than 1.  For example,
the corrections are expected to have an isospin dependence which permits corrections to be tested with enhanced
sensitivity in $I=2$ systems\cite{TownerHardy09}.  Recent branching ratio measurements in $^{32}$Ar
decay and planned measurements in $^{20}$Mg, $^{24}$Si, $^{28}$S, $^{32}$Ar, $^{36}$Ca and $^{40}$Ti should
provide further insight into the reliablity of the isospin-mixing calculations in use at present.

\subsection{Neutron Decay}
$I=1/2$ decays between isobaric analog states provide a reasonable alternative to superallowed decays,
and an effective cross-check of some of the theoretical corrections implicit in the analysis of nuclear decays.
From a nuclear structure point of view, the simplest such process is neutron decay.  Extraction of $V_{ud}$ from
neutron $\beta$-decay does not require the application of corrections for isospin-symmetry breaking effects,
$\delta_C$, or nuclear-structure effects, $\delta_\mathrm{NS}$. However, it should be noted that the transition-dependent
radiative correction, $\delta R^\prime$,  and the nucleus-independent radiative correction, $\Delta_R^V$, are
still needed. In contrast to nuclear $\beta$-decays between $0^+$ states, which sample only the weak vector interaction,
neutron $\beta$-decay proceeds via a mixture of the weak vector and axial-vector couplings. Consequently, three
parameters are required for a description of neutron $\beta$-decay: $G_F$, the fundamental weak interaction constant;
$\lambda=g_A/g_V$, the ratio of the weak axial-vector and vector coupling constants; and the parameter of interest,
$V_{ud}$.  The axial coupling constant, $g_A$, is of fundamental interest\cite{Abel08}, in that it plays an important
role in a variety of particle physics and astrophysics problems.  Although no theoretical methods exist to
calculate values for $g_A$ which are competitive with the accuracy of the experimental value at present, this quantity
remains a central target for high precision lattice calculations, with some technical progress on this problem
reported recently\cite{juttner-CKM2010}.

The neutron lifetime, $\tau_n$, can be written in terms of the above parameters as \cite{MarcianoSirlin06,Czarneki04,Abel08}:
\begin{equation}
\frac{1}{\tau_n} = \frac{G^2_Fm^5_e}{2\pi^3}\left|V_{ud}\right|^2\left(1+3\lambda^2\right)f(1+\mathrm{RC}),
\end{equation}
where $f = 1.6887\pm0.00015$ is a phase space factor, which includes the Fermi function contribution\cite{Wilkinson82},
and $(1 + \mathrm{RC}) = 1.03886(39)$ denotes the total effect of all electroweak radiative
corrections\cite{MarcianoSirlin06,Czarneki04}. Therefore:
\begin{equation}
\left|V_{ud}\right|^2 = \frac{4908.7(1.9)~\mathrm{s}}{\tau_n\left(1+3\lambda^2\right)}.
\label{neutron:vud}
\end{equation}

\begin{figure}
\centering
\includegraphics[height=6.0 cm]{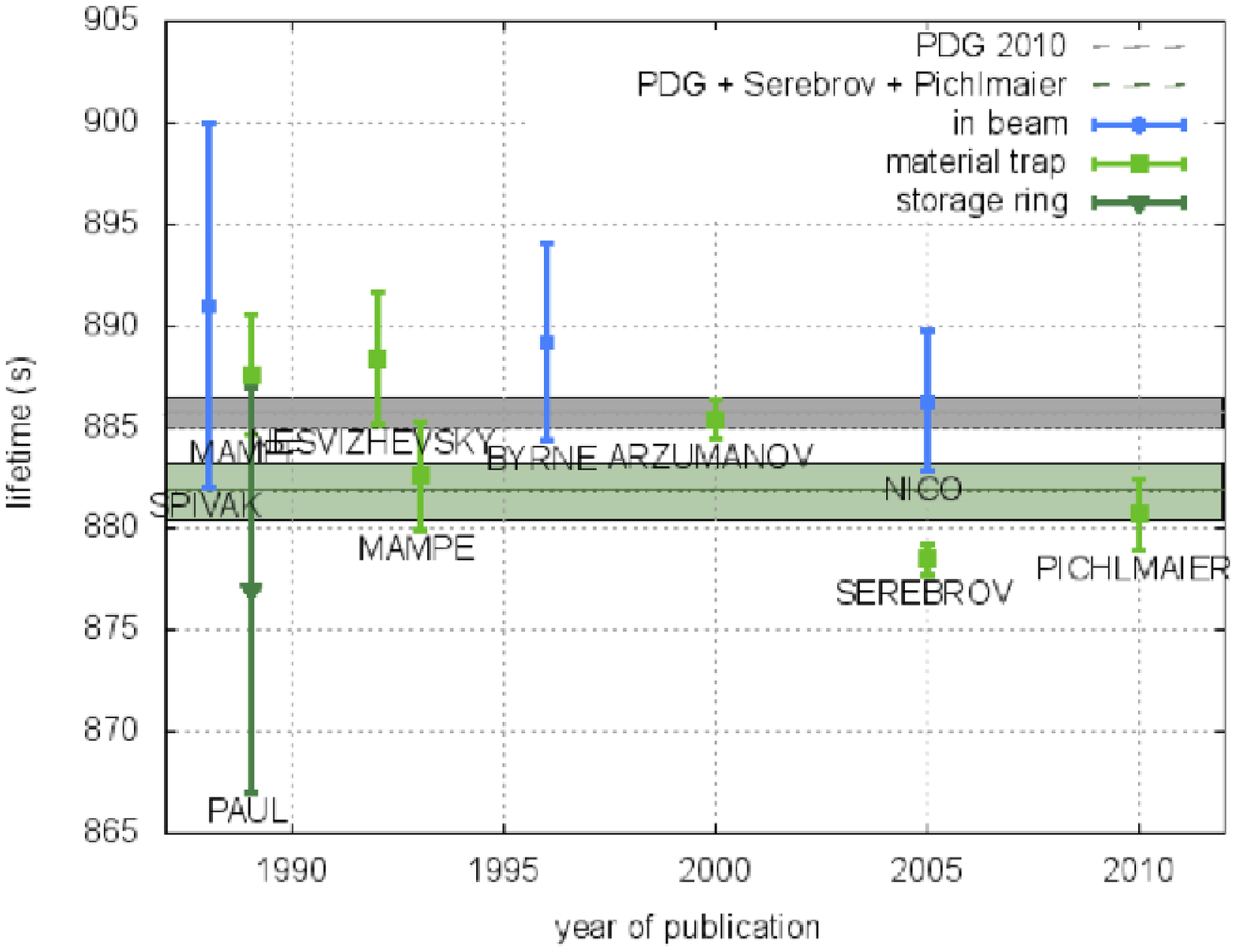}
\includegraphics[height=6.0 cm]{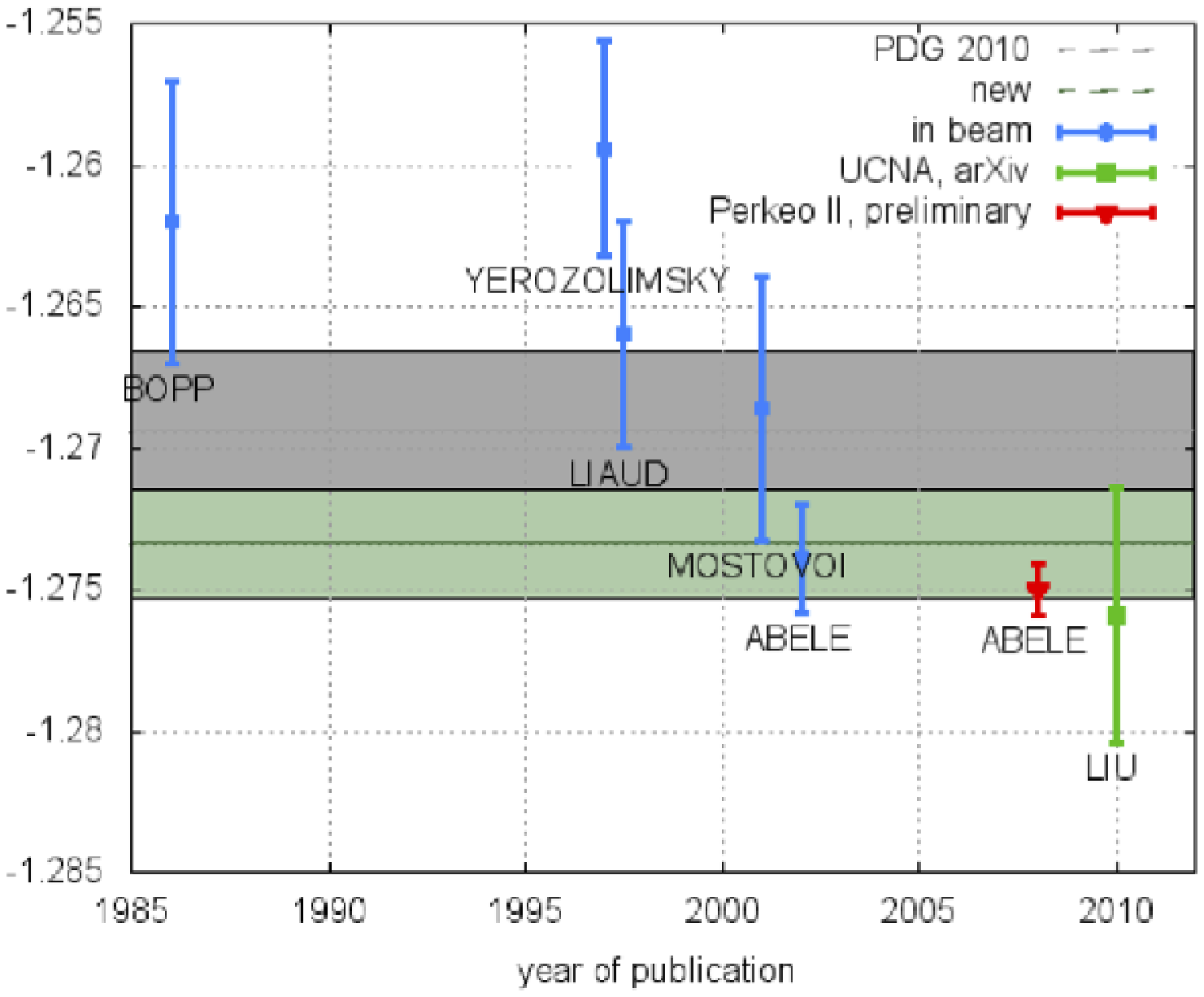}
\caption{Pictures from~\cite{markisch-CKM2010}. Left: measurements of the neutron lifetime from 1988 to 2011. Right: measurements of
the $\lambda = g_A /g_V$ in neutron decay from 1985 to 2011.  Figure taken from M{\" a}rkisch\cite{markisch-CKM2010}.}
\label{fig:taun}
\end{figure}

The present experimental status of neutron lifetime is not crystal clear. A precise measurement from 2005 by
Serebrov lies $\sim6.5$ standard deviations away from the average of the others (Fig.~\ref{fig:taun}, left).
In its 2008 revision, the PDG excluded it adding that the resulting average is ``suspect.''\cite{PDG08}.

B. M{\" a}rkisch\cite{markisch-CKM2010} reviewed the status of in-beam experiments with cold neutrons
(CN, kinetic energies of $\sim3$~meV) and storage experiments with
ultra-cold neutrons (UCN, kinetic energies of $<\simeq 300$~neV). In the first scheme, the number $n_\beta$ of beam neutrons
decaying in a defined acceptance (a space length ${\ell}$) are counted and this provides an absolute measurement of
$\tau_n$: $n_\beta=-N_0/\tau_n~\mathrm{exp}(-\ell /v_n\tau_n)$, with $v_n$ the mean velocity of the neutron beam.
In the second scheme, $N_1$ UCN are trapped in a given volume (``bottle'' or material trap) which is opened after a given
time $\Delta T$. The number of UCN flowing out, $N_2$, provides a relative measurement of $\tau_n$: in the absence of loss
mechanisms other than decay, $\tau_n = \Delta T~\mathrm{ln}(N_1/N_2)$. The evaluation of losses due to interaction with the trap
walls is particularly delicate and often constitutes the dominant systematic error.

A new report of a UCN lifetime measurement appeared in 2010 by the MAMBO II collaboration~\cite{mambo2_2010},
in which the trap volume was varied simultaneously with the storage time $\Delta T$ so that the effect of wall
interaction losses is minimized when comparing different neutron countings. The result, $\tau_n=880.7(1.3)_{stat}(1.2)_{syst}$~s
is below the 2010 PDG average by 2.5 standard deviations\cite{PDG2010} and reinforces the need to resolve discrepancies in
treatment of systematic errors in these experiments (Fig.~\ref{fig:taun}, left).  Given the existence of several
experiments now with statistically significant discrepancies towards smaller values of the liftime, the MAMBO II
collaboration pointed out that a reasonable strategy moving forward is to expand the error bars on the global value
to be consistent with the scatter in reported values when incorporating all present experimental data.  The result of such a survey
is $\tau_n = 881.8(1.4)$, where the uncertainties were scaled by a factor of 2.7 following PDG rules.

A value for the ratio $\lambda$ of axial-vector to vector current contributions to neutron decay can be directly extracted
from measurements of correlation coefficients in polarized neutron $\beta$-decay. Neglecting recoil corrections,
the correlation coefficients $a$ (the $e$-$\nu_e$ asymmetry), $A$ (the $\beta$-asymmetry), and $B$ (the $\nu_e$-asymmetry)
can be expressed in terms of $\lambda$ (to leading order) as \cite{Wilkinson82,GardnerZhang01}:
\begin{equation}
  a=\frac{1-\vert \lambda \vert ^2}{1+3\vert \lambda \vert ^2}\mbox{, }A=-2\frac{\vert \lambda \vert^2+{\rm Re}(\lambda )}
{1+3\vert \lambda \vert^2}\mbox{, }B=2\frac{\vert \lambda \vert ^2-{\rm Re}(\lambda )}{1+3\vert \lambda \vert^2}.
\end{equation}
In the standard model, the complex part of $\lambda$ is expected to be negligible\cite{Abel08,Herc97}. The PDG 2010 values for these correlation
parameters are $a = −0.103(4)$, $A = −0.1173(13)$, and $B = 0.983(4)$.  Although $B$ has been measured
to the highest precision (0.41\%), the sensitivity of $B$ to $\lambda$ is a factor $\sim10$ less than that of $a$ and $A$.
The neutron $\beta$-asymmetry yields the most precise result for $\lambda$. The experimental results are shown in Fig.~\ref{fig:taun}, right.
Until 2009, all reported angular correlation measurements in neutron decay have utilized cold neutron beams.  Recently, the UCNA experiment
utilized UCN to improve the systematic error on $A$, at present at the level of 1~\%\cite{Liu10}.  When we include the result of this
measurement and a preliminary value for the PERKEO II experiment\cite{Abel08}, the value of axial form factor is
$\lambda = -1.2734(19)$, where the errors have been scaled by 2.3 to be consistent with the experimental scatter.

As reported by M{\" a}rkisch\cite{markisch-CKM2010}, if we utilize the most recent values for the neutron lifetime and the
axial form factor, we can use Eq. \ref{neutron:vud} to extract a value of $V_{ud} = 0.9743(2)_{RC}(8)_{\tau _n}(12)_{\lambda}$
from neutron decays.

\subsection{Mirror Decays}
Transitions between isobaric analog states with nuclear isospin $1/2$ provide a second alternative to superallowed
decays.  Like neutron decay, these decays proceed through both vector and axial-vector couplings, and therefore
two measurements must be performed on these systems to fix the value of the axial-vector coupling and $V_{ud}$.
The equation used to determine $V_{ud}$ from these decays is\cite{melconian-CKM2010}:
\begin{equation}
\vert V_{ud} \vert ^2 = \frac{5831.3(2.3)}{Ft^{mirror}(1 + \frac{f_A}{f_V}\rho ^2)},
\end{equation}
where $f_A/f_V$ is the ratio of statistical rate functions for axial/vector currents and $\rho=C_A M_{GT} /C_V M_V$ is the ratio
of Gamow-Teller to Fermi strengths for the decay.  As with the neutron, a correlation typically determines the ratio $\rho$.
Recent work\cite{naviliatcuncic} analyzed the published data for these decays, identifying five transitions for which
both the $Ft$ value and at least one correlation parameter is measured, to permit an extraction of $V_{ud}$. These
are $^{19}$Ne, $^{21}$Na, $^{29}$P, $^{35}$Ar and $^{37}$K. Taken together, they determine $V_{ud}=0.9719(18)$.
Recent measurements of the lifetimes of $^{19}$Ne, $^{21}$Na and $^{37}$K should produce modest improvements, but for
significant reduction of the uncertainties, improved angular correlations measurements are required for these systems.

\subsection{Pion Beta-Decay}
$V_{ud}$ can also be obtained from the pion $\beta$-decay, $\pi^+\to\pi^0\nu e^+(\gamma)$, which is a pure vector
transition between two spin-zero members of an isospin triplet  and is therefore analogous to the superallowed
nuclear decays. Like neutron decay, it has the advantage that there are no nuclear-structure dependent corrections to be
applied. Its major disadvantage, however, is that the pion has a very small branching ratio to this channel,
${\cal O}(10^{-8})$. Using the branching ratio measured by the PIBETA group\cite{Poca04}, a value of $V_{ud}=0.9742(26)$
is determined from pion decay.

\subsection{$V_{ud}$ Summary}
\begin{figure}
\centering
\includegraphics[height=8.0cm]{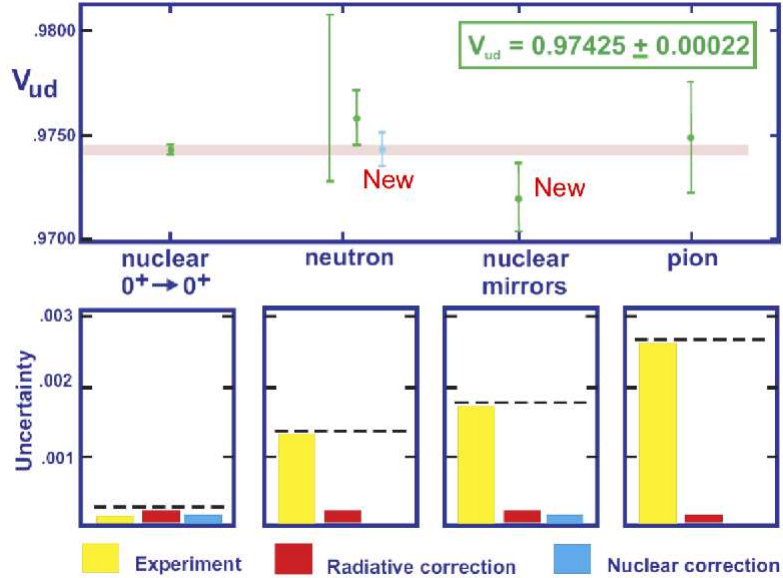}
\caption{Summary from~\cite{towner-CKM2010}. Top: $V_{ud}$ average values from superallowed Fermi nuclear $\beta$ decays, from
neutron decays, from nuclear decays of ``mirror'' nuclei, and from $\pi^+\to\pi^0\nu e$ decays. Bottom: error
components for each method are disentangled.}
\label{fig:vud}
\end{figure}

The status of $V_{ud}$ is summarized in Fig.~\ref{fig:vud} from I. Towner's contribution in this conference\cite{towner-CKM2010}.
The values extracted from all four decay systems: $I=0$ superallowed decays, neutron decay, $I=1/2$ mirror decays and pion 
decay are consistent with each other, with the value extracted from superallowed decays still significantly more
precise than the values determined from other decays.

\section{Status of $V_{us}$}
\label{sec:vus}
Three main sources are used to extract $V_{us}$. Sorted according to the present uncertainty, these are: the measurement of the
transition widths for semileptonic kaon decays, for leptonic kaon decays, and the spectral analysis
of inclusive semileptonic $\tau$ decays to strange hadrons, $\tau\to X_s l \nu$. For the first two, the
most important recent improvements are related to the uncertainty of theoretical inputs from lattice calculations.
The total uncertainty for $V_{us}$ from $K$ decays is around 0.5\%. Novel methods have been proposed for $\tau$ analysis,
which might reach interesting levels of accuracy in the near future.

\subsection{$V_{us}$ from $K$ decays}

\subsubsection{Semileptonic Kaon decays}

The photon-inclusive $K\to\pi{\mathit l}\nu$ decay rates are conveniently decomposed as~\cite{PDG08}:
\begin{equation}
\Gamma(K{\mathit l}3(\gamma))= \frac{G_F M_K^5}{192\pi^3} C_K S_\mathrm{EW} \left|V_{us} f_+^{K^0\pi^+}(0)\right|^2
I_K^{\mathit l}(\lambda_+,\lambda_0)\left(1 + \delta_\mathrm{EM}^{K{\mathit l}}+\delta_{SU(2)}^{K\pi}\right),
\end{equation}
where $C_K^2 = 1 (1/2)$ for the neutral (charged) kaon decays, $S_\mathrm{EW}$ is the short distance
electroweak correction, $f_+^{K^0\pi^+}(0)$ is the $K\to\pi$ vector form factor at zero momentum
transfer, and $I_K^{\mathit l}(\lambda_+,\lambda_0)$ is the phase space integral which depends on the 
(experimentally accessible) slopes of the form factors (generically denoted by $\lambda_{+,0}$).
Finally, $\delta_\mathrm{EM}^{K {\mathit l}}$ represent EM
channel-dependent long distance radiative corrections and $\delta K_\mathrm{SU(2)}^{K\pi}$ is a correction induced
by strong isospin breaking.

Measurements of BR's exist for five transitions, $K_S\to\pi\mu\nu$ being the only one still missing, while the slopes $\lambda$ are
extracted from $K_L$ and $K^\pm$ modes.  Experimental inputs from $K_L$ (KLOE, KTeV, NA48), $K^\pm$ (E865, KLOE, NA48, ISTRA+),
and $K_S$ (KLOE, KTeV, NA48) provide an average value of $V_{us} f_+^{K^0\pi^+}(0) = 0.2163\pm0.0005 $, 
and a consistent picture, with a $\chi^2=0.77$ for four degrees of freedom. For details, see the talk by B. Sciascia for
the FlaviaNet working group~\cite{Sciascia10}.  The average is performed taking the correlation among various inputs into account
and after a critical and careful analysis of the the literature, in particular a consistent treatment of radiative corrections, which
are at the few percent level, and for the $K^+$ lifetime measurements. Therefore, the set used for the above average
slightly differs from that of the PDG group~\cite{PDG2010}. The results for each mode, and the contributions to the uncertainty are
listed in Table~\ref{tab:vusf0}.  This shows that for the charged modes theory is still driving the error, due to the $SU(2)$-breaking
correction $\delta_{SU(2)}^{K\pi}=2.9\pm0.4\%$. 
\begin{table}
\begin{tabular}{c|c|c|c|c|c|c}
Mode  & $V_{us} f_+^{K^0\pi^+}(0)$ & Error, \% & \multicolumn{3}{c|}{Error source, \%} \\
            &              &        &  BR    &      $\tau$      & $\delta_\mathrm{EM}^{K{\mathit l}}+\delta_{SU(2)}^{K\pi}$ & $I_K^{\mathit l}(\lambda_+,\lambda_0)$  \\ \hline
$K_{Le3}$    & 0.2163(3)    &  0.26  &  0.09  &     {\bf 0.20}  &  0.11     & 0.06 \\ \hline   
$K_{L\mu3}$  & 0.2166(6)    &  0.29  &  0.15  &     {\bf 0.18}  &  0.11     & 0.08 \\ \hline   
$K_{Se3}$    & 0.2155(13)   &  0.61  &  {\bf 0.60}  & 0.03  &  0.11     & 0.06 \\ \hline   
$K^\pm_{e3}$ & 0.2160(11)   &  0.52  &  {\bf 0.31}  & 0.09  &  {\bf 0.40} & 0.06 \\ \hline   
$K^\pm_{\mu3}$ & 0.2158(14)  & 0.63  &  {\bf 0.47}  & 0.08  &  {\bf 0.39} & 0.08 \\ \hline   
\end{tabular}
\caption{Results for $V_{us}f_+^{K^0\pi^+}(0)$ from each mode from the FlaviaNet working group~\cite{Sciascia10}. 
The total uncertainty is disentangled and the contributions in bold are considered leading.}
\label{tab:vusf0}
\end{table}

The kinematical dependence 
of the form factors is usually parametrized either on the basis of a systematic mathematical expansion or by imposing additional physical constraints, such as the presence of 
a pole corresponding to a single resonance or via dispersion relations. In the last case, a QCD prediction of the form factor at the non-physical Callan-Treiman point allows
a consistency check of the form factor measurement and the lattice inputs (vector and scalar form factors at zero momentum transfer).
A new preliminary result for form-factor slopes from analysis of
$K^\pm_{\mu3}$ decays from NA48/2 has been presented by M. Veltri~\cite{Veltri10}, 
which seems to solve previous tensions between results from NA48 and the other experiments for the scalar form factor.
NA62, a future experiment located in the same $K$ beam line used by NA48, is under construction and will take data in 2014. Data taking with an improved $K^+$ beam and 
the NA48 detector has been performed by the newly formed NA62 collaboration in 2007. 
Data under analysis are expected to increase by a factor of 10 the statistics of selected $K^+_{e3,\mu3}$ with respect to NA48/2, thus allowing 
significant improvements on the BRs and uncertainty of $K^+_{e3,\mu3}$ decays.

As shown by R. Escribano~\cite{Escribano2010}, 
the $K\pi$ form factors can be constrained complementing $K_{l3}$ data with
$\tau\to K\pi\nu_\tau$ data via a dispersive representation. Significant improvements can be derived in terms of accuracy of the form factor slopes. 
In particular, the total
error on the slopes from muon modes can be reduced by a factor of $\sim0.6$.

After a precise measurement of the $K_S$ lifetime in 2010, KLOE is the only experiment providing results for all of the experimental quantities involved. 
The average performed with the same 
technique as above but using only inputs from KLOE has the same accuracy as the world average, $f_+(0)V_{us}=0.2157\pm0.0006$, with $\chi^2=7$ for four degrees of freedom.
Detector and machine upgrades foreseen in the next few years (the so-called KLOE-2 program) 
are expected to improve on the above figure, as presented by E. De Lucia~\cite{DeLucia10}. In particular, after 5~fb$^{-1}$ of additional integrated luminosities,
the total uncertainties on $K_L$, $K^\pm$, and $K_S$ are expected to be reduced to 0.3\%, 0.1\%, and 0.03\%. Moreover, the BR for $K_{Se3}$ will be measured at 0.6\%, 
$K_{S\mu3}$ will be measured for the first time, with an uncertainty of $\sim2\%$, and the errors on $K_L$ BR's will be reduced by a factor of three. After these efforts,
the uncertainty for the world average of $f_+(0)V_{us}$ should be reduced by a factor of 2, to the level of 0.15\%.

\begin{figure}
\centering
\includegraphics[height=8.0cm]{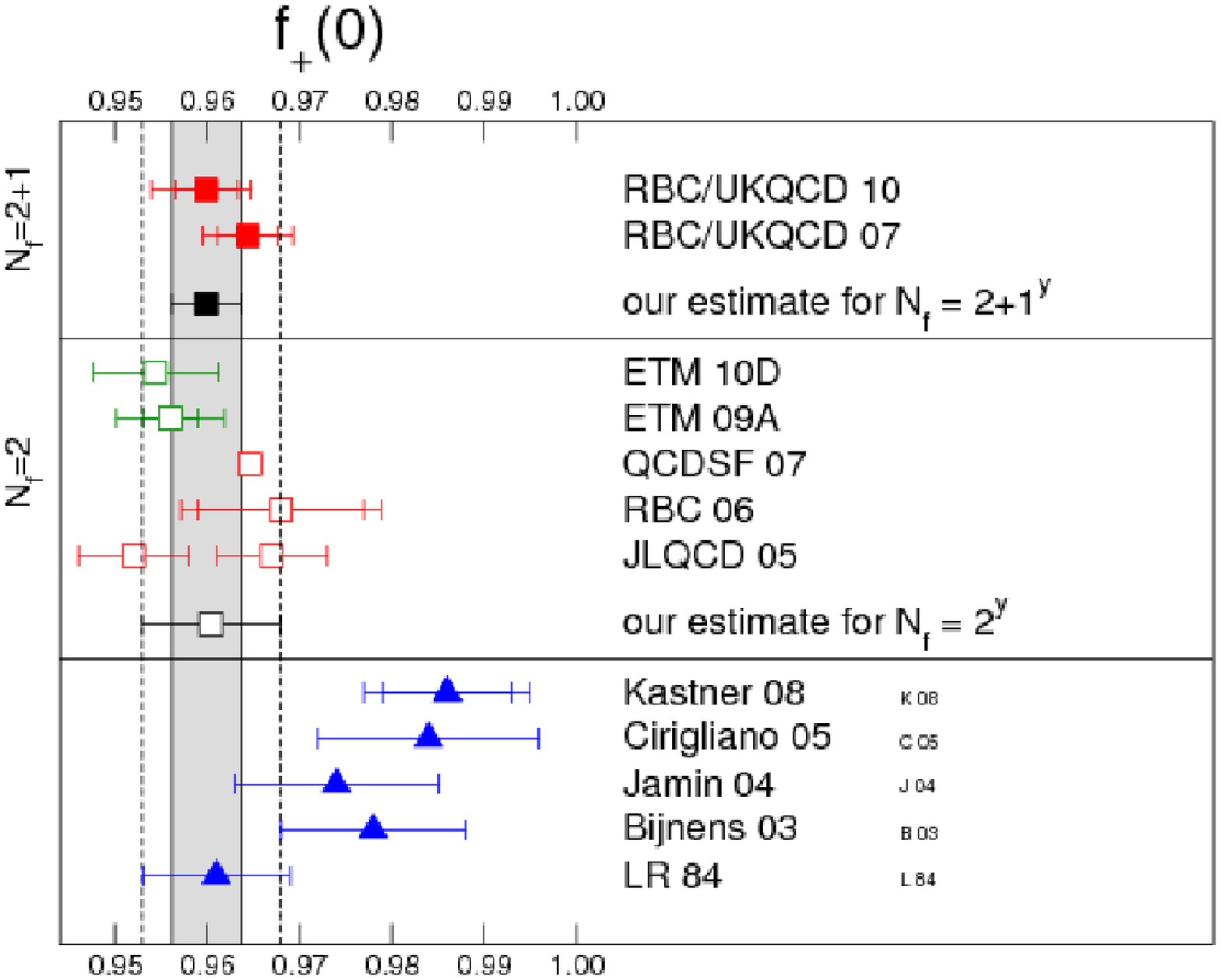}
\includegraphics[height=8.0cm]{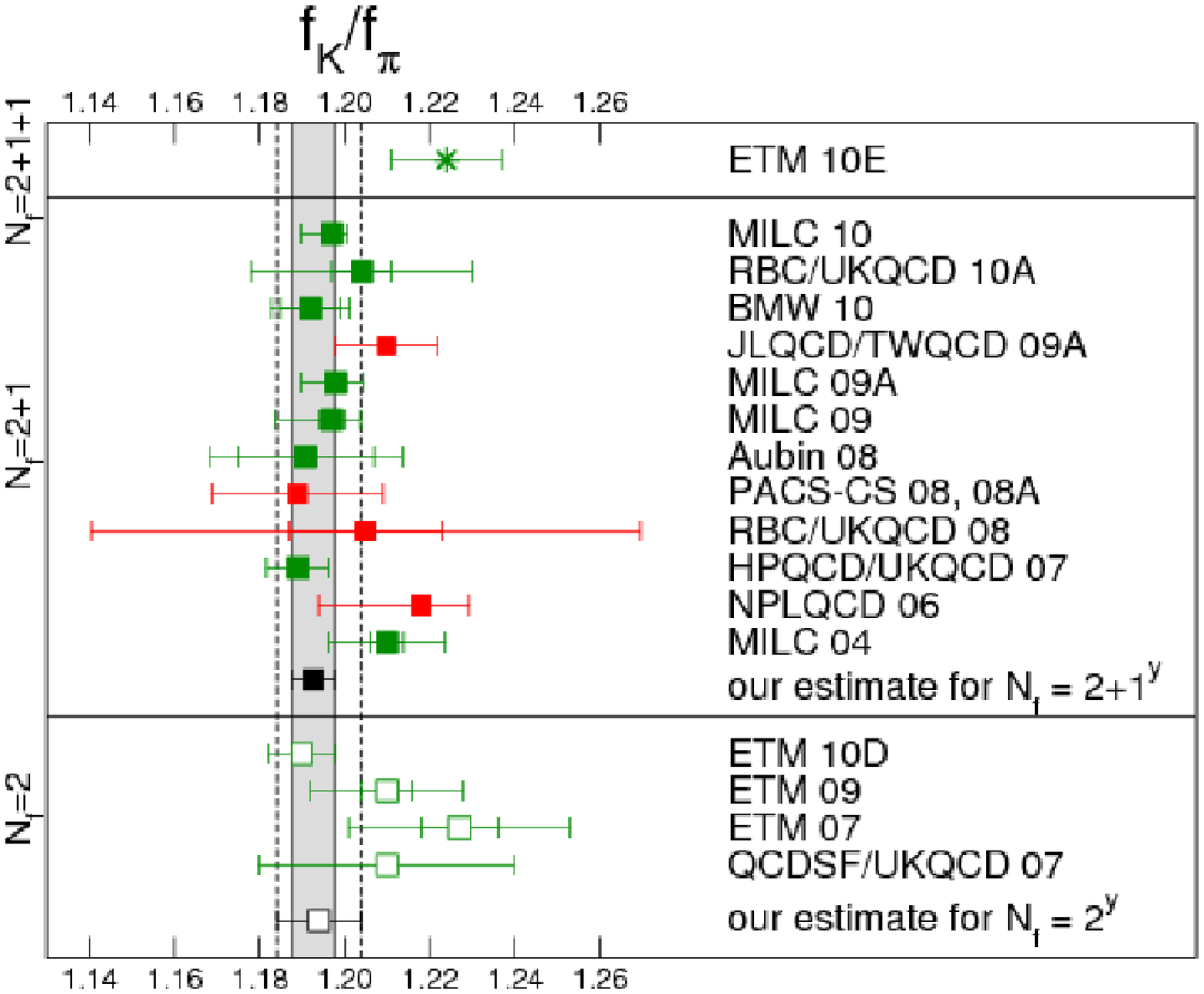}
\caption{Plots from~\cite{Juttner10}. Results for the $K^0\to\pi^+$ form factor at zero momentum transfer $f_+(0)$ (top) and 
for the ratio $f_K/f_\pi$ of scalar form factors (bottom) 
from simulations with $2+1$- (filled squares) and 2-flavors (open squares) and from other techniques
(filled triangles).}
\label{fig:lattice}
\end{figure}

In recent years, lattice techniques to calculate $f_+(0)$ reached an unprecedented precision. A representation of results in literature is shown in the top plot of Fig.~\ref{fig:lattice}. Studies from a
specific group within the FlaviaNet framework, called Flavianet Lattice Averaging Group (FLAG) 
led to a systematic characterization of the lattice results from the various groups, as discussed in the talk by A. Juttner~\cite{Juttner10}.
Each technique used is classified by FLAG according to the accuracy of the chiral extrapolation, the errors from effects due to the finite volume 
of the lattice used and from the extrapolation to the continuum, 
and finally the publication status. A total precision of 0.5\% on $f_+(0)$ is within reach. At present, two state-of-the-art results are available. These are 
quoted distinguishing between actions with $2+1$-active flavors,
\begin{equation}
f_+(0) = 0.9599(34)_\mathrm{stat}(^{+31}_{-47})_{\chi\mathrm{extrap}}(14)_\mathrm{continuum}\mbox{ RBC/UKQCD coll. 2010~\cite{RBCUKQCD10f0} },
\end{equation}
and 2-active flavors ($N_f$),
\begin{equation}
f_+(0) = 0.9560(57)_\mathrm{stat}(35)_{\chi\mathrm{extrap}}(37)_\mathrm{fin.size,cont.}(28)_\mathrm{quench.}\mbox{ ETM coll. 2009~\cite{ETM09f0}}.
\end{equation}
The effect of the dynamical strange quark is not visible at the current level of precision for this observable.
A reduction of the pion mass in the lattice evaluation is mandatory to reduce systematic errors due to chiral extrapolation.
As the lattice spacing is further reduced, the estimate of the statistical error will be more and more delicate, due to possible auto-correlations which are
not easily taken into account. 

Using the world-average value from FlaviaNet, $f_+(0)=0.959(5)$, the value of $V_{us}$ is:
\begin{equation}
\label{eq:vusfinal}
|V_{us}|=0.2254(13).
\end{equation}

\subsubsection{Leptonic Kaon decays}

The other main experimental input relevant for $V_{us}$ is the ratio of leptonic widths for $K\to\mu\nu(\gamma)$ and $\pi\to\mu\nu(\gamma)$:
\begin{equation}
\frac{\Gamma(K_{\mu2(\gamma)})}{\Gamma(\pi_{\mu2(\gamma)})} = \frac{\left|V_{us}\right|^2}{\left|V_{ud}\right|^2}\times\frac{\left|f_K\right|^2}{\left|f_\pi\right|^2}\times
\frac{M_K(1-m_\mu^2/M_K^2)^2}{M_\pi(1-m_\mu^2/M_\pi^2)^2}\times(1+\delta_{em}),
\end{equation}
where long-distance electromagnetic corrections are precisely evaluated as $\delta_{em}=-0.0070\pm0.0018$ 
and the ratio of kaon and pion decay constants $f_K/f_\pi$ is a theoretical input from
lattice calculations. On the experimental side, no major improvements have 
been recently achieved and the information has an uncertainty at the level of
0.2~\%: $f_K/f_\pi|V_{us}/V_{ud}|=0.27599(59)$. 

As for $f_+(0)$, in recent years lattice techniques improved dramatically for the evaluation of $f_K/f_\pi$, too. 
A summary of results in literature is shown in the bottom panel of Fig.~\ref{fig:lattice}. 
Very good agreement is shown among various evaluations and no contributions from sea strange quarks are visible at the current precision of data. 
The averages by the FLAG group are again quoted distinguishing between actions with $2+1$-active flavors,
\begin{equation}
\label{eq:fkfp}
f_K/f_\pi = 1.193(6)\mbox{ BMW 10~\cite{BMW10fkfp}, MILC 09~\cite{MILC09fkfp}, and HPQCD/UKQCD 08~\cite{HPQDC08fkfp} },
\end{equation}
and 2-active flavors ($N_f$),
\begin{equation}
f_K/f_\pi = 1.210(18)\mbox{ ETM Coll. 2009~\cite{ETM09fkfp}}.
\end{equation}
A reduction of the pion mass in the lattice evaluation is within reach and this will allow a reduction of systematic errors due to chiral extrapolation.
As the lattice spacing is further reduced, the estimate of the statistical error will be more and more delicate, due to possible auto-correlations which are
not easily taken into account. On the same subject, A. Ramos reported in detail on the evaluation of $f_K/f_\pi$ by the BMW Collaboration, with $2+1$ flavors. 
The extrapolation to the physical point
has been performed both with $SU(2)$ or $SU(3)$ chiral perturbation theory and with a Taylor expansion. The average for all of the approaches is evaluated weighting for the
fit quality and the difference of results from each method has been taken as systematic error. A similar procedure is also applied for the evaluation of the error due to 
finite volume effects, to the
extrapolation to the continuum, and to the contribution of excited states. The result is:
\begin{equation}
f_K/f_\pi = 1.192(7)_\mathrm{stat}(6)_\mathrm{syst}\mbox{ [BMW Coll. 2010]~\cite{BMW10fkfp}},
\end{equation}
where the systematic error is dominated by chiral extrapolation (0.0045) and by the extrapolation to the continuum (0.0033). To further reduce the uncertainty due to
chiral extrapolation, new simulation configurations have been generated 
such that the pion and kaon masses are equal to the physical masses~\cite{Hoelbing10}. 
Therefore, future results from the BMW collaboration are expected to improve significantly on the total error
for $f_K/f_\pi$.

Using the value from Eq~\ref{eq:fkfp}, the result for $V_{us}/V_{ud}$ is:
\begin{equation}
\label{eq:vusvud}
\left|\frac{V_{us}}{V_{ud}}\right|=0.2312(13).
\end{equation}


\subsection{$V_{us}$ from $\tau$ decays}
The spectral analysis
of inclusive semileptonic $\tau$ decays to strange hadrons, $\tau\to X_s l \nu$, involves both vector and axial vector transitions. The inclusiveness allows use of the operator product expansion (OPE) and other analytical constraints. The presence of
a hard scale (the $\tau$ mass $m_\tau$) is beneficial in terms of accuracy.

The measured invariant mass squared distribution $s$ of the final state hadrons allows determination of the corresponding moments
\begin{equation}
R_\tau^{k\mathit l}  = \int_0^{m_\tau^2}\mathrm{d}s\left(1-\frac{s}{m_\tau^2}\right)^k\left(\frac{s}{m_\tau^2}\right)^{\mathit l}\frac{\mathrm{d}R_\tau}{\mathrm{d}s},
\end{equation}
where the rate $R_\tau$ for $\tau\to X_s l \nu$ is normalized to that for $\tau\to e\overline{\nu}_e \nu_\tau$. Provided the rates are measured distinguishing among
Cabibbo-suppressed (S) or Cabibbo-allowed (NS) transitions, one can assess the following quantity:
\begin{equation}
\label{eq:vustau}
\delta R_\tau^{k\mathit l} = \frac{R_{\tau,\mathrm{NS}}^{k\mathit l}}{V_{ud}} - \frac{R_{\tau,\mathrm{S}}^{k\mathit l}}{V_{us}},
\end{equation}
which vanishes in the $SU(3)$ symmetry limit and allows a precise theoretical prediction through OPE. In particular, $\delta R_\tau$ involves matrix elements for
not-strange ($ud$ final states, for $\tau\to X\ell \nu$) or strange ($us$ final states, for $\tau\to X_s\ell \nu$) operators of dimension $D\geq 2$, each with a suppression factor $\sim (m_s/m_\tau)^D$. 
An accuracy of 10\% on the $SU(3)$-breaking contribution 
translates into an error of 0.4\% on $V_{us}$.

The highest sensitivity to $V_{us}$ from Eq.~\ref{eq:vustau} stems from the $k,\mathit l=0,0$ momentum. In order to evaluate $V_{us}$,
one has to use a value for the strange mass from other sources
as an input so that the theoretical expectation for $\delta R_{\tau}^{k \mathit l}$ can be assessed. In a recent compilation~\cite{gamiz08}, the theoretical 
expectation reads:
\begin{equation}
\label{eq:vustau_theo}
\delta R_\tau^{00} = 0.227(54)\mbox{ with }m_s(\mathrm{2~GeV}) = 95(20)\mathrm{~MeV}\mbox{ and }m_s(m_\tau)=100(10)\mathrm{~MeV}.
\end{equation}
After a new set of results from the $B$ factories for $\tau$ branching fractions, the experimental findings are $R_{\tau,\mathrm{NS}}^{00}=3.478(11)$ and 
$R_{\tau,\mathrm{S}}^{00}=0.1617(40)$. The experimental data set after completion of ALEPH analyses in 2005 was $R_{\tau,\mathrm{NS}}^{00}=3.469(14)$ and $R_{\tau,\mathrm{S}}^{00}=0.1677(50)$. The significantly lower value of $R_{\tau,\mathrm{S}}$ pushes down the value of $V_{us}$.
Using $V_{ud}=0.97425(22)$ from superallowed decays 
(Eq.~\ref{eq:vudsa}), one gets 
\begin{equation}
V_{us}=0.2165\pm0.0026_\mathrm{exp}\pm0.0005_\mathrm{th},
\end{equation}
As shown by K. Maltman in his contribution to CKM 2010~\cite{Maltman10}, 
sizeable changes of the experimental determination of $R_{\tau,\mathrm{S}}$ are to be expected from the full analysis of the Babar and Belle data samples. 
In particular, the high-multiplicity decay modes are not well known at present and their effect has been just roughly estimated or simply ignored. 
Thus, the result above and its present discrepancy with respect to the $V_{us}$ determination from $K$ decays might change dramatically in the near future, 
after modes such as $K^-\pi^0\pi^0,$ $K3\pi,$ $K4\pi,$ and others are addressed by Babar and Belle.
However, it is important to underline that the final error of the $V_{us}$ determination 
from $\tau$ decay will probably be dominated by the experimental uncertainties. 
If $R_{\tau,\mathrm{S}}$ were known with a 1\% uncertainty, the resulting $V_{us}$ uncertainty would be reduced to around 0.6\%, i.e. $\pm0.0013,$ 
making $\tau$ decay the best source of information about $V_{us}$. 

\section{New Physics tests from $K$ decays}

\subsubsection{$CP$ violation}

Within the SM, violations of $CP$ are described by a single phase in the CKM matrix. In this picture, the two $CP$-violating quantities
for the $K^0$ and the $B^0$ systems, $\epsilon$ and $\sin(2\beta)$ are necessarily related. Tests of this relation can be sensitive to new-physics contributions.

The $K_S$ and $K_L$ eigenstates of the equation of motion are written in terms of
$K^0$ and $\overline{K^0}$ states as:
\begin{equation}
\left|K_{S,L}\right\rangle = \frac{(1+\overline{\epsilon})|K^0\rangle \pm (1-\overline{\epsilon})|\overline{K^0}\rangle}{\sqrt{2(1+|\overline{\epsilon}|^2)}}.
\end{equation}
Provided the quantity $\langle K_S | K_L \rangle = 2\mathrm{Re}(\overline{\epsilon})/(1+|\overline{\epsilon}^2|)$ differs from zero, $CP$ violation is implied. In terms of the
mass and width matrices $M_{ij}$ and $\Gamma_{ij}$, 
\begin{equation}
\mathrm{Re}(\overline{\epsilon}) = \mathrm{sin}(\phi_\epsilon)\mathrm{cos}(\phi_\epsilon)
\left(
\frac{\mathrm{Im}(M_{12})}{\Delta M_K}-\frac{\mathrm{Im}(\Gamma_{12})}{2\mathrm{Re}(\Gamma_{12})}
\right),
\end{equation}
where $\mathrm{tan}(\phi_\epsilon) = 2 \Delta M_K/\Delta \Gamma$ can be assessed from the $K_S$--$K_L$ mass and width differences. The ratios $\eta_{\pm,00}$ of $K_L$ to $K_S$
transition amplitudes to a pair of pions $\pi^+\pi^-$, $\pi^0\pi^0$ are the experimental inputs for the evaluation of $\overline{\epsilon}$:
\begin{equation}
\epsilon_K \equiv \frac{2\eta_{+-}+\eta_{00}}{3} = e^{i\phi_\epsilon}\mathrm{sin}(\phi_\epsilon)\left(\frac{\mathrm{Im}(M_{12})}{\Delta M_K} + \xi \right),
\end{equation}
where $\xi = \mathrm{Im}(a_0)/\mathrm{Re}(a_0)$ and $a_0 e^{i\delta_0}$ parametrizes the transition of $K^0$ to an isospin-zero $\pi\pi$ final state. From experiment, $\epsilon_K$
is known at the 1\% level, $\epsilon_K=2.228(11)\times10^{-3}$. The theoretical uncertainties from the right hand side arise from $M_{12}$ and $\xi$. As summarized in the talk
by D. Guadagnoli~\cite{Guadagnoli10}, before 2008 it was customary to approximate $\xi=0$ and $\phi_\epsilon=45^\circ$ and to consider only the short distance contribution to $M_{12}$:
\begin{equation}
M_{12}^{SD} \propto \left\langle\overline{K^0}\right|\left(\overline{s}\gamma_\mu^L d\right)\left(\overline{s}\gamma^{\mu L}d\right)\left|K^0\right\rangle.
\label{eq:m12}
\end{equation}
These approximations were reasonable, given the uncertainty of more than 10\% for the evaluation of $M_{12}^{SD}$, dominated by the uncertainty on $B_K$:
\begin{equation}
|\epsilon_K| = k_\epsilon C_\epsilon B_K A^2\overline{\eta}\left(-\eta_1 S_0(x_c)(1-\lambda^2/2) + \eta_3 S_0(x_c,x_t) + \eta_2 S_0(x_t)A^2\lambda^2(1-\overline{\rho})
 \right),
\label{eq:epsilonk}
\end{equation}
where $A\equiv |V_{cb}|/\lambda^2$ and $\lambda=|V_{us}|$; $6\sqrt{2}\pi^2 C_\epsilon=(G_F^2 f_K^2 m_K M_W^2)/\Delta M_K$; $S_0$ are the Inami-Lin functions depending on 
$x_{c,t}=m^2_{c,t}/M_W^2$, parametrizing the charm-top contributions to the box diagrams; $\eta_{1,2,3}$ contain the corresponding short-distance QCD contributions evaluated
from perturbative calculations; $\overline{\rho}$ and $\overline{\eta}$ are the
normalized coordinates of the unitarity triangle vertex; finally, the bag parameter $B_K$ parametrizes the difference  
from the $\Delta S=2$ amplitude of Eq.~\ref{eq:m12} and that evaluated as a product of two $\Delta S=1$ amplitudes, using the vacuum-insertion approximation. $B_K$ is 
calculated from lattice techniques.

\begin{figure}
\centering
\includegraphics[height=8.0cm]{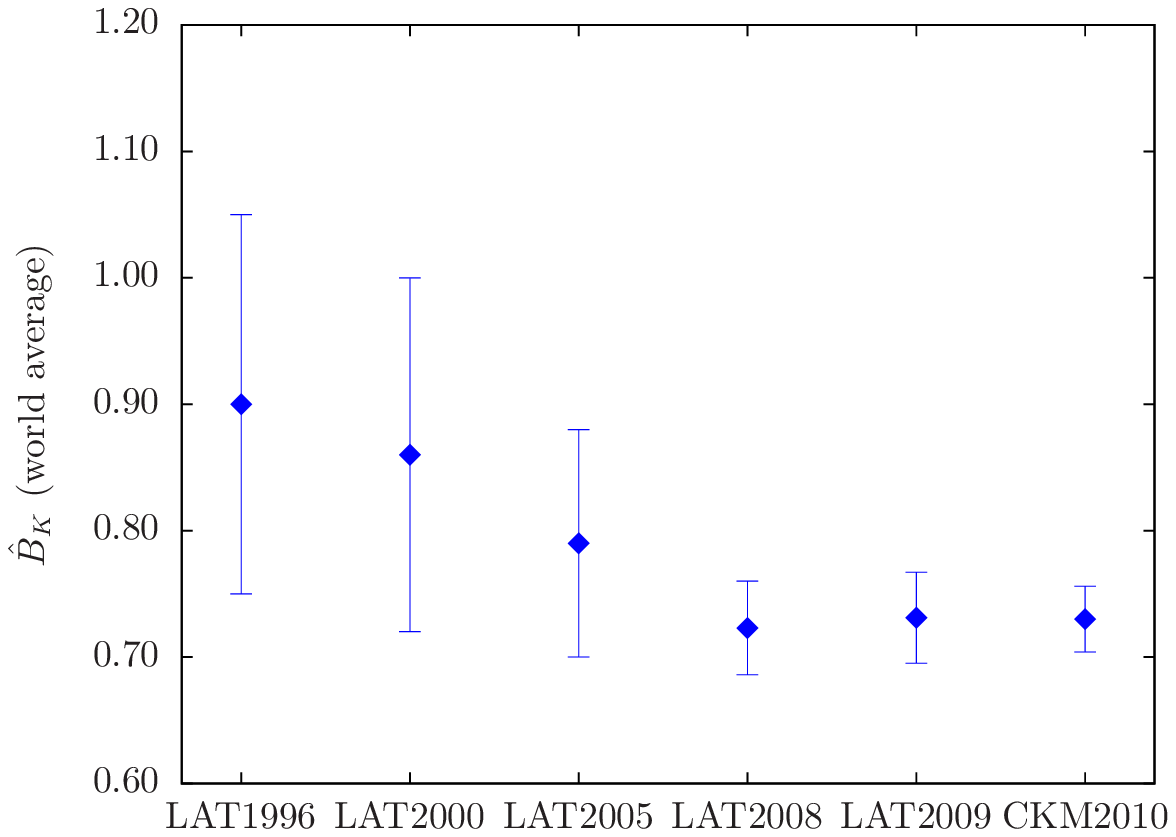}
\caption{Results for the world average of the bag parameter $B_K$, from~\cite{Dimopoulos10}. Lattice results, filled dots, can be compared 
with the prediction from unitarity triangle fit, $B_K=0.87(8)$.}
\label{fig:epsilon}
\end{figure}

Recent significant improvements have been achieved for $B_K$: the accuracy has been reduced from 18\% to 4\% in the last 10 years or so, see Fig.~\ref{fig:epsilon}. 
This progress is the due to the use of unquenched simulations with $N_f=2$ and $N_f=2+1$ dynamical quarks,
with high statistics and better control of discretization errors and extrapolation to the physical point. For details,
see the report by P. Dimopulos in this conference~\cite{Dimopoulos10}. 

As shown by D. Guadagnoli~\cite{Guadagnoli10}, given the present accuracy on $B_K$ the previous approximations are not valid. Given the values of $\Delta M$ and $\Delta\Gamma$, 
the measured value of the phase $\phi_\epsilon$ is 43.5$^\circ$.
The value of $\xi$ can be evaluated exploiting the
dominance of the transition amplitudes with isospin variation $\Delta I=1/2$ over those with $\Delta I=3/2$ and relating the $\Delta I=1/2$ part of $\xi$ to the direct $CP$
violation parameter $\epsilon^\prime/\epsilon_K$. Moreover, the long-distance contributions to $\mathrm{Im}(M_{12})$ are evaluated in chiral perturbation theory.
In summary, these corrections amount to a multiplicative factor $k_\epsilon=0.94\pm0.02$. 

At present, the leading uncertainty in the constraint in the $\overline{\rho}$--$\overline{\eta}$ plane is due to the value of $V_{cb}$, which enters with the fourth power
in Eq.~\ref{eq:epsilonk}: $\delta(V_{cb})/V_{cb}\sim2.5\%$ implies $\delta(\epsilon_K)/\epsilon_K\sim10\%$. The error on the $B_K$ parameter accounts 
for an additional 5\%. For comparison, the uncertainty from inputs of $B$ physics, the relative error on $\overline{\eta}(1-\overline{\rho})$, is $\sim8\%$.
The UTfit group quotes a fit tension of $\sim2.6\sigma$ from the comparison of $CP$ violation in $K$ and $B$ systems~\cite{UTfit10}.

\subsubsection{Lepton universality tests in $K_{\ell 2}$ decays}

The decay $K^\pm\!\to e^\pm\nu$ is strongly suppressed by a factor of
$\sim$~few$\times$10$^{-5}$, because of conservation of angular
momentum and the vector structure of the charged weak current. It
therefore offers the possibility of detecting minute contributions from
physics beyond the SM. This is particularly true of the
ratio $R_K=\Gamma(\ke)/\Gamma(\km)$ which, in the SM, is
calculable without hadronic uncertainties \cite{marci,Cirigliano:2007xi}.

 It has been pointed out that in a supersymmetric framework 
sizable violations of  lepton universality can be expected
 in $K_{l2}$ decays~\cite{paride}. At the tree level, 
 lepton flavor violating terms are forbidden in the MSSM. 
 However, these appear at the one-loop level, where an effective 
 $H^+ l \nu_\tau$ Yukawa interaction is generated. Following the
 notation of Ref.~\cite{paride},
 the non-SM contribution to $R_K$ can be written as 
\begin{equation}
R_K^{LFV} \approx R_K^{SM} \left[ 1 + \left(
\frac{m_K^4}{M_{H^\pm}^4} \right) \left( \frac{m_\tau^2}{m_e^2} \right) |\Delta_{13}|^2 \tan^6 \beta \right]~.
\label{eqn:susy}
\end{equation}
 The lepton flavor violating coupling $\Delta_{13}$, being generated at the 
loop level, could reach values of $\cO(10^{-3})$.
 For moderately large $\tan \beta$ values, 
 this contribution may therefore
 enhance $R_K$ by up to a few percent.
 Since the additional term in Eq.~\ref{eqn:susy} goes with the forth power
 of the meson mass, no sizeable effect is expected for $\pi_{l2}$ decays.

$R_K$ is {\em defined} to be inclusive of inner bremsstrahlung (IB), 
ignoring however structure-dependent direct emission
(DE) contributions.  A calculation
including order $e^2p^4$ corrections in chiral perturbation
theory gives~\cite{Cirigliano:2007xi}:
\begin{equation}
\label{eq:rksm}
 R_K = (2.477\pm0.001)\times 10^{-5}.
\end{equation}
$R_K$ is not directly measurable, since
IB cannot be distinguished from DE on an event-by-event basis.
Therefore, in order to compare data with the SM prediction at the
percent level or better, the DE contribution must be carefully estimated and subtracted.\footnote{
 The same arguments
 apply in principle to $\Gamma$(\km). However, helicity
 suppression is weaker in this case. IB must be included and DE can be safely
 neglected.}

 $R_K$ has been measured very recently by KLOE~\cite{RKKLOE} and NA62~\cite{RKNA62} on samples
 of about 14,000 \ke\ events and of about 60,000 \ke\ events respectively. For the NA62 measurement, based on the analysis of 40\% of the total data set, see the contribution by C. Lazzeroni in this 
conference~\cite{Lazzeroni10}.
 KLOE also performed a study of the photon spectrum in \kedg.
 Both analyses are inclusive of the IB contribution, while the DE has been
 treated differently.

 KLOE define the rate $R_{10}$ as:
 \begin{equation}
 R_{10}=\Gamma(\ke(\gamma),\ E_\gamma<10\mathrm{\ MeV})/\Gamma(\km).
\label{eq:R10}
\end{equation}
Evaluating the IB spectrum to ${\mathcal O}(\alpha_{\rm em})$
with resummation of leading logarithms, $R_{10}$ includes
$93.57\pm0.07\%$ of the IB,
\begin{equation}
R_{10}=R_K\times(0.9357\pm0.0007).
\label{eq:R10new}
\end{equation}
The DE contribution in this range is expected to be negligible.
 However, due to resolution effects on the measurement of $E_\gamma$, the event sample used by KLOE to measure
$R_{10}$  still contains a small DE contribution,
 in particular for decays with high electron momentum in the CM, $p_e$.
In order to subtract this contribution, 
KLOE has also measured the differential width
\begin{equation}
\frac{\mathrm{d}R_\gamma}{\mathrm{d}E_\gamma}=
\frac{1}{\Gamma(\km)}\frac{\mathrm{d}\Gamma(\ke\gamma)}{\mathrm{d}E_\gamma},
\label{eq:rgamma}
\end{equation}
 for $E_\gamma\!>10$ MeV and $p_e\!>200$ MeV
requiring photon detection, both to test ChPT predictions for the
DE terms and to reduce possible systematic uncertainties on the
$R_{10}$ measurement.

The DE contribution is strongly rejected (by about a factor of 10)
in the NA62 analysis by vetoing non-collinear photons.
The residual DE is evaluated and subtracted using the KLOE result.

Different approaches have been also employed in discriminating 
\ke\ events from the \km\ background ($\sim10^5$ times larger).
 The $e/\mu$ separation with calorimeters is more effective at
 high energy. Therefore, the resulting background rejection factor
 is about 50 times larger for NA62 than for KLOE.
 This gap is partly recovered by KLOE exploiting the 
 better kinematical rejection, due to the low momenta involved.
 The total resulting effective background contamination is about
 14\% in KLOE and $(8.8\pm0.3)\%$ in NA62.

\begin{figure}[t]
\centering
\includegraphics[width=0.45\textwidth]{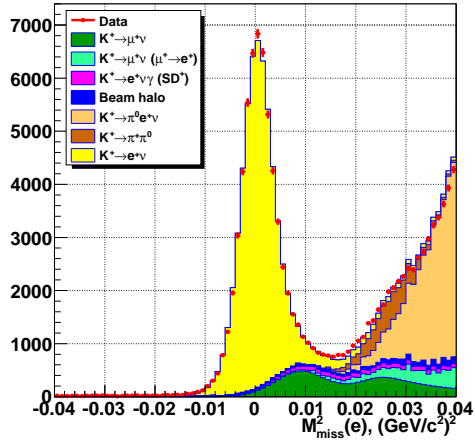}
\caption{\label{fig:ke2NA48}
NA62 $K\to e\nu$ selection: $m_{miss}^2$ distribution, for data (dots) and MC signal (solid line). The various background contributions are represented 
by filled histograms. }
\end{figure}

From the kaon and decay particle momenta, 
${\bf p}_K$ and ${\bf p}_{\rm d}$, the squared mass
$m_\ell^2$ of the lepton for the decay
 $K\to\ell\nu$ assuming zero missing mass or the squared missing mass
 $m_{miss}^2$ assuming the electron mass for the decay particle can be computed.
 The distribution of $m_{miss}^2$ is shown in 
 Fig.\ref{fig:ke2NA48} for the NA62 data.

The NA62 result for $R_K$ has been checked as a function of the lepton laboratory momentum, comparing 10 independent momentum bins.
A $\chi^2$ fit to the measurements of $R_K$ in each momentum bin has been performed, taking into account 
the bin-to-bin correlations between the systematic errors. 
To validate the assigned systematic uncertainties, extensive stability checks have been performed in bins of 
kinematic variables and by varying selection criteria and analysis procedures.

The final results from KLOE and NA62 are listed in Tab.~\ref{tab:ke2kmu2}. The 0.5\% uncertainty of the result from the 
NA62 experiment is still dominated by the statistical error, $\sim0.4\%$, thus leaving room for future improvements when the analysis
of the entire data will be completed.
Combining these recent results with the current PDG value, the new world average is:
\begin{equation}
R_K  = ( 2.487 \pm 0.012 ) \times 10^{-5}.
\label{eqn:ke2kmu2}
\end{equation}
This is in good agreement with the SM expectation of Eq.~\ref{eq:rksm} and,
with a relative error of $0.5\%$, it improves by an order of magnitude with respect to the 2008 world average.

\begin{table}[t]
\small
  \begin{center}
      \begin{tabular}{lc}
        \hline \hline
        & $R_K$ $[10^{-5}]$  \\ \hline
        PDG 2008         & $2.45 \pm 0.11$ \\
        NA62             & $2.487 \pm 0.013$ \\
        KLOE             & $2.493 \pm 0.031 $ \\ \hline
        SM prediction    & $2.477 \pm 0.001$ \\
        \hline \hline
                                                  & \\*[-3mm]
      \end{tabular}
      \caption{Results and prediction for $R_K$.}
      \label{tab:ke2kmu2}
  \end{center}
\end{table}

\begin{figure}[t]
\centering
\resizebox{0.45\textwidth}{!}{\includegraphics{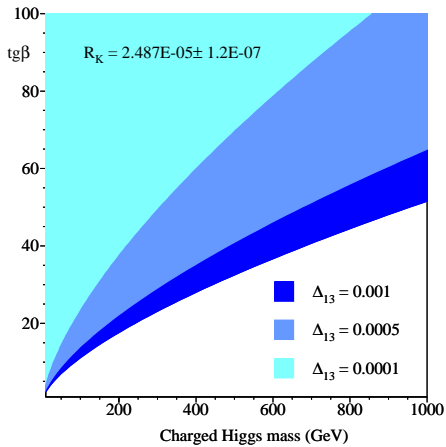}}
\caption{Exclusion limits at $95\%$ CL on $\tan \beta$ and the charged
Higgs mass $M_{H^\pm}$ 
from the new world average of $R_K$, for different
values of $\Delta_{13}$. }
\label{fig:susylimit}
\end{figure}

 The world average result for $R_K$ gives strong constraints 
 for $\tan \beta$ and $M_{H^\pm}$, as shown in Fig.~\ref{fig:susylimit}.
 For values of $\Delta_{13} \approx 5 \times 10^{-4}$
 and  $\tan \beta > 50$ the charged Higgs mass is pushed 
 above 1000~GeV/$c^2$ at 95\% CL.

\section{Status of the unitarity test}
\label{sec:end}
After fitting the measurements of $V_{ud}$ from Eq.~\ref{eq:vudsa}, $V_{us}$ from Eq.~\ref{eq:vusfinal}, 
and of $V_{us}/V_{ud}$ from Eq.~\ref{eq:vusvud}, the
following values are extracted:
\begin{equation}
\label{eq:vusandvud}
|V_{ud}|=0.97425(22)\mbox{ and }|V_{us}|=0.2253(9),
\end{equation}
with a $\chi^2=0.014$ for one degree of freedom. The unitarity constraint is satisfied:
\begin{equation}
\label{eq:unitarity}
|V_{ud}|^2+|V_{us}|^2+|V_{ub}|^2 = 0.9999(4)_{V_{ud}}(4)_{V_{us}}.
\end{equation}
As shown by B. Marciano~\cite{marciano-CKM2010}, this is a clear proof of the conserved vector current assumption and of the radiative corrections predicted by the SM.
From this result, one might derive a prediction for the $Z$ mass, $M_Z=90(7)$~GeV. 
Many constraints for different new physics scenarios can be inferred,
probing effective scales roughly of the order of 10~TeV for tree-level transitions.


\begin{thebibliography}{99}
\bibitem{PDG2010} K. Nakamura {\it et al.} (Particle Data Group), ``J. Phys. {\bf G 37}, 075021 (2010) and partial update for the 2012 edition
(neutron particle listing).
\bibitem{debevec-CKM2010} P. Debevec for the MuLan Collaboration, CKM 2010 Proceedings. For the latest result on $\mu^+$ lifetime,
see D.~M.~Webber {\it et al.} [ MuLan Collaboration ], ``Measurement of the Positive Muon Lifetime and Determination of the Fermi
Constant to Part-per-Million Precision,'' Phys.\ Rev.\ Lett.\  {\bf 106}, 041803 (2011).
\bibitem{marciano} S.M. Berman, Phys. Rev. 112, 267 (1958); T. Kinoshita and A. Sirlin, Phys. Rev. 113, 1652 (1959); S.M. Berman
and A. Sirlin, Ann. Phys. 20, 20 (1962). W.J. Marciano, Phys. Rev. D 60, 093006 (1999).
\bibitem{Pak08}A. Pak and A. Czarneki, Phys. Rev. Lett. 100, 241807 (2008).
\bibitem{TownerHardy09} J.~C.~Hardy, I.~S.~Towner,``Superallowed $0^+\to 0^+$ nuclear beta decays: A New survey with precision
tests of the conserved vector current hypothesis and the standard model,'' Phys.\ Rev.\  {\bf C79}, 055502 (2009).
\bibitem{melconian-CKM2010}D. Melconian, CKM2010 proceedings, ``Status of Nuclear Beta-Decay Measurements,'' arXiv:1108.2530 (2011).
\bibitem{marciano-CKM2010} W. Marciano, CKM2010 proceedings, ``Standard model and beta-decay theory,''
\url{http://indico.cern.ch/contributionDisplay.py?sessionId=3&contribId=9&confId=96378} (2010).
\bibitem{MarcianoSirlin06} W. J. Marciano, A. Sirlin, ``Improved calculation of electroweak radiative corrections and the value of Vud,''
Phys. Rev. Lett. 96 (2006) 032002.
\bibitem{towner-CKM2010}I. Towner, CKM2010 proceedings, ``Theoretical Treatment of Superallowed Nuclear Decays,'' ArXiv:1108.2516 (2011).
\url{http://indico.cern.ch/contributionDisplay.py?sessionId=3&contribId=10&confId=96378} (2010).  For the latest treatment of the
nuclear corrections, see I. Towner and J. Hardy, ``Comparative tests of isospin-symmetry-breaking corrections to $0^+ \to 0^+$ nuclear
beta decay,'' Phys. Rev. C {\bf 82}, 065501 (2010).
\bibitem{Czarneki04} A. Czarneki, W. J. Marciano, A. Sirlin, ''Precision measurements and CKM unitarity,'' Phys. Rev. D {\bf 70}, 09300 (2004).
\bibitem{Abel08} H. Abele, ``The neutron. Its properties and basic interactions,'' Prog. Part. Nucl. Phys. {\bf 60}, 1-81 (2008).
\bibitem{Wilkinson82} D. H. Wilkinson, ''Analysis of neutron beta decay,'' Nucl. Phys. {\bf A377}, 474-504 (1982).
\bibitem{juttner-CKM2010} A. Juttner, ``Lattice nucleon matrix elements: $g_A$,''
\url{http://indico.cern.ch/contributionDisplay.py?sessionId=3&contribId=113&confId=96378} (2010). 
\bibitem{PDG08} C. Amsler {\it et al.} (Particle Data Group), Phys. Lett. {\bf B667}, 1 (2008).
\bibitem{markisch-CKM2010} B. M{\" a}rkisch, CKM 2010 Proceedings, ``Experimental Status of $V_{ud}$ from Neutron Decay,'' ArXiv:1107.3422 (2010).
\bibitem{mambo2_2010} A. Pichlmaier {\it et al.} (MAMBO-II Collaboration), ``Neutron lifetime measurement with the UCN trap-in-trap
MAMBO II,'' Phys. Lett. {\bf B693} (2010), 221-226.
\bibitem{Herc97}P. Herczeg and I. B. Khriplovich, ``Time-reversal violation in $\beta$-decay in the standard model,'' Phys.
Rev. D {\bf 56}, 80 (1997).
\bibitem{GardnerZhang01} S. Gardner, C. Zhang, ''Sharpening low-energy, standard-model tests via correlation coefficients in neutron
beta-decay'', Phys. Rev. Lett. {\bf 86} (2001) 5666–5669.
\bibitem{Liu10}J. Liu {\it et al.}, Phys. Rev. Lett. {\bf 105}, 181803 (2010).
\bibitem{naviliatcuncic} O.~Naviliat-Cuncic, N.~Severijns, ``Test of the Conserved Vector Current Hypothesis in T=1/2 Mirror Transitions
and New Determination of $|Vud|$,'' Phys.\ Rev.\ Lett.\  {\bf 102}, 142302 (2009).
\bibitem{Poca04}D. Pocanic {\it et al.}, ``Precise measurement of the $\pi ^+ \rightarrow \pi ^0 e^+ \nu$ branching ratio,''Phys. Rev.
Lett. {\bf 93}, 181803 (2004).

\bibitem{Sciascia10} B.~Sciascia for the FlaviaNet Kaon Working Group Collaboration, CKM2010 proceedings, ``$V_{us}$ and precise Standard Model tests,''
  arXiv:1101.5024 (2011).
\bibitem{Veltri10} M.~Veltri, for the NA48/2 and NA62 Collaborations,
``$K^{\pm}_{\mu3}$ Form Factors Measurement at NA48/2,'' arXiv:1101.5031 (2011).
\bibitem{Escribano2010} D.~R.~Boito, R.~Escribano, M.~Jamin,
  ``The $K\pi$ vector form factor and constraints from $K_{l3}$ decays,''
  arXiv:1101.2887 (2011).
\bibitem{DeLucia10} E. De Lucia for the KLOE and KLOE-2 Collaborations, 
CKM2010 proceedings, ``Determination of $V_{us}$ at the KLOE experiment: Present results and future perspectives,''
arXiv:1101.5016 (2011).
\bibitem{Juttner10} A. Juttner, to appear in the proceedings of SPIRES Conference C11/06/13, e-Print: arXiv:1109.1388 [hep-ph].

\bibitem{RBCUKQCD10f0} P. A. Boyle et al., (RBC/UKQCD Collaboration), e-Print: arXiv:1004.0886 [hep-lat].
\bibitem{ETM09f0} V. Lubicz et al., (ETM Collaboration), Phys. Rev. {\bf D 80}, 111502 (2009), e-Print: arXiv:0906.4728 [hep-lat].
\bibitem{Ramos10} A. Ramos, {\it et al.} (BMW Collaboration), ``$F_K/F_\pi$ from the Budapest-Marseille-Wuppertal Collaboration,''
arXiv:1101.3968 (2011).
\bibitem{BMW10fkfp} S. Durr et al., (BMW Collaboration), Phys. Rev. {\bf D 81}, 054507 (2010), e-Print: arXiv:1001.4692 [hep-lat]. 
\bibitem{MILC09fkfp} A. Bazavov et al. (MILC Collaboration), PoS {\bf LAT2009}, 079 (2009), e-Print: arXiv:0910.3618 [hep-lat].
\bibitem{HPQDC08fkfp} E. Follana et al., (HPQCD/UKQCD Collaboration), Phys. Rev. Lett. {\bf 100}, 062002 (2008), e-Print: arXiv:0706.1726 [hep-lat].
\bibitem{ETM09fkfp} B. Blossier et al., (ETM Collaboration), JHEP {\bf 0907}, 043 (2009), e-Print: arXiv:0904.0954 [hep-lat].
\bibitem{Hoelbing10} C.~Hoelbling,
  ``Light hadron spectroscopy and pseudoscalar decay constants,''
  PoS {\bf LATTICE2010}, 011 (2010), 
  arXiv:1102.0410 (2011).

\bibitem{gamiz08} E. Gamiz, {\it et al.}, ``Theoretical progress on the Vus determination from $\tau$ decays'',
arXiv:0709.0282 (2007).

\bibitem{Maltman10} K. Maltman, ``Hadronic $\tau$ Decay Based Determinations of $|V(us)|$'', arXiv:1101.1985 (2011).

\bibitem{Guadagnoli10}
  D.~Guadagnoli,
  ``On the consistency between CP violation in the K vs. Bd systems within the Standard Model,''
  arXiv:1102.2760 (2011).

\bibitem{Dimopoulos10} P.~Dimopoulos,
 ``$K^0$--$\bar{K}^0$ on the Lattice,''
  arXiv:1101.3069 (2011).

\bibitem{UTfit10}
  A. Bevan, {\it et al.} (UTfit Collaboration), arXiv:1010.5089 (2010).

\bibitem{marci} W.J. Marciano and A. Sirlin, Phys. Rev. Lett. {\bf 71} (1993) 3629; 
  M. Finkemeier, Phys. Lett. B {\bf 387} (1996) 391.

\bibitem{Cirigliano:2007xi} V. Cirigliano and I. Rosell,
  Phys. Rev. Lett.  {\bf 99} (2007) 231801.

 \bibitem{paride}
  A.~Masiero, P.~Paradisi and R.~Petronzio,
  Phys.\ Rev.\  D {\bf 74} (2006) 011701
  [arXiv:hep-ph/0511289].

\bibitem{RKKLOE}
{Spadaro, T., \textit{et al.} \textup{(the KLOE collaboration)}}.
\newblock {Precise measurement of $BR(K\to e\nu(\gamma))/B(K\to \mu\nu(\gamma))$ and study of $K\to e\nu\gamma$}.
\newblock  Eur.\ Phys.\ J.\  C {\bf 64} (2009) 627
\newblock  [Erratum-ibid.\  {\bf 65} (2010) 703]
\newblock  [arXiv:0907.3594 [hep-ex]].

\bibitem{RKNA62}
  C.~Lazzeroni {\it et al.} [ NA62 Collaboration ],
  \newblock {Test of Lepton Flavour Universality in $K^+ \to l^+\nu$ Decays}.
  \newblock Phys.\ Lett.\  {\bf B698 } (2011)  105-114.
  \newblock [arXiv:1101.4805 [hep-ex]].

\bibitem{Lazzeroni10} C. Lazzeroni (for the NA62 collaboration),
``Lepton universality tests with leptonic kaon decays,''  
  arXiv:1101.5817 (2011).

\end{thebibliography}
\end{document}